\documentstyle[preprint,aps,prd,eqsecnum]{revtex}
\begin{document}
\draft
\tighten
\preprint{DO-TH-99/06}
\title{\bf A hierarchy of sum-rules in out of equilibrium QFT}
\author{\bf Julien F. J. Salgado
        \footnote{e-mail:salgado@yoda.physik.uni-dortmund.de}}
\address
{Institut f\"ur Physik,
Universit\"at Dortmund\\
D-44221 Dortmund,
Germany}
\date{\today}
\maketitle
\begin{abstract}
Generalising a result of classical mechanics an infinite set of conserved
quantities can be found for the bare equations of motion describing the
evolution of a scalar field in out of equilibrium quantum field theory,
in the large $N$ approximation, with initial conditions corresponding to
a thermal system of the free Hamiltonian.
Using these new conserved quantities, sum-rules relating integrals over the
mode-functions (momenta) can be derived.
More, the corresponding renormalised quantities can also be computed out
thus giving information about the evolution of the already known
renormalised equations; finally it is also possible to write a renormalised
version of the sum-rules.
\end{abstract}

\section*{Introduction}

Nowadays, out of equilibrium quantum field theory is believed to act as a
powerful tool
in order to understand systems within which the energy or
particle density are high. This is typically the case for cosmological
problems
\cite{Baacke:1997rs,Boyanovsky:1997cr,Cormier:1998wk}
such as inflation
\cite{Boyanovsky:1997ae,Boyanovsky:1997mq,Boyanovsky:1998xt,Boyanovsky:1996ks,Cormier:1998nt},
for which the study of preheating \cite{Baacke:1997kj,Boyanovsky:1996sq}
is now well understood with a self consistent treatment in out of
equilibrium QFT. It is now understood how the particle production behaves
\cite{Boyanovsky:1998cr,Calzetta:1989vs}, in particular what
is the contribution of the quantum fluctuation in this process.
The physics of heavy ion can also be studied \cite{Boyanovsky:1998yp} with
this theory in particular the hadronisation process of the quark gluon
plasma. 
The analysis of initial condition is also well understood
\cite{Baacke:1998zz,Boyanovsky:1998ba}.

During the last couple of years many impressive results have been found
using accurate numerical computations with different approximations in the
form of the equation of motion: one loop
\cite{Baacke:1997se},
Hartree, leading order in large
$N$ expansion and even recently beyond the leading order. 
The variety was not only in the methods but also in the models and the
physical problems that were studied: systems with fermions \cite{Baacke:1998di},
electrodynamics \cite{Boyanovsky:1998aa,Cooper:1989kf}, non-Abelian gauge
fields \cite{Baacke:1997kj},...

Although the scope of the theory is becoming wider and wider there only few
analytical results compared to the numerical works. The new analytical
result are mainly well controlled expansion \cite{Boyanovsky:1997cr}, that have been occasionally
re-summed
\cite{Boyanovsky:1998aa,deVega:1997yw}
.
Notwithstanding all those attempts exact results have seldom been
found \cite{Alamoudi:1998pt}, only two sum-rules were found
\cite{Boyanovsky:1998zg} in the case of a scalar field with initial
conditions corresponding to a zero temperature bath, one has to
mention that a generalisation for non zero temperature has been predicted
using one loop computation \cite{Baacke:1998zy}.

The aim of the present paper is to prove the existence of a hierarchy of
sum-rules that generalise the previous result. Using such sum-rules one
should be able to find more on the system, since just using the first to
sum-rules one was able to find out the asymptotical equation of state of the
quantum system.

In order to get a progressive understanding the classical results will be
summarised in the following section. Using those result it will be possible
to construct in section \ref{section_bare} the bare  conserved quantities
for a scalar field in out of equilibrium field theory. Bare sum-rules
expressed in terms of momenta, {\it i.e.} integrals over the modes and
their derivates, will be explicited in section \ref{section_sum}.
The analysis of the renormalisation process will be described in section
\ref{section_renormalisation} for the conserved quantities; in
section \ref{section_renormalised_momenta} it will be shown how the momenta
and the sum-rules can be renormalised in a such way that the sum-rules
still hold for renormalised quantities.

\section{Classical oscillator in radial quartic potential}

We just recall some results on the radial quartic potential, that have been
studied by by Choodnovsky and Choodnovsky \cite{CC1,CC2} and extended by
Wojciechowski \cite{SW}. We shall give a brief review of the results of
Grosse \cite{Grosse}, we consider a N-dimensional system whose evolution
is given from the following Hamiltonian:
\begin{equation}
H=\frac12\sum_{i=1}^N (p_i^2+r_kq_i^2)
        +\frac14\Biggl(\sum_{i=1}^Nq_i^2\Biggr)^2,
\end{equation}
where all the masses are different.

One can say that the $O(N)$ invariance is broken by different masses.
Nonetheless the system remains integrable, since one can
find a complete set of (independent) and commutating conserved quantities
which can be written as:
\begin{equation}
\label{Kclassical}
K_i=\sum_{i\ne j}\frac{(q_ip_j-q_jp_i)^2}{r_j-r_i}
        +2p_i^2+2r_iq_i^2+q_i^2\sum_{j=1}^Nq_j^2
\end{equation}
It is straight forward to check the following:
\begin{equation}
\Bigr\{K_i,K_j\Bigl\}=0
\end{equation}

One should notice that, the Hamiltonian can easily be reconstructed from the
$K_i$, one trivially has:
\begin{equation}
\label{Eclassical}
H=\frac14\sum_{i=1}^N K_i
\end{equation}

\section{Bare conserved quantities in out of equilibrium QFT}
\label{section_bare}
The main idea is to generalise the previous result to the infinite number of
equation describing the motion of a scalar field in out of equilibrium
quantum field theory. In the leading order in the large $N$ expansion the
equations of motion read (for a detailed derivation one should refer to
\cite{Baacke:1998zy,Cooper:1994hr,Cooper:1997ii,Cormier:1998wk})
\begin{eqnarray}
\Biggl(\frac{d^2}{dt^2}+{\cal M}(t)^2\Biggr)\phi(t)&=&0\nonumber\\
\Biggl(\frac{d^2}{dt^2}+k^2+{\cal M}(t)^2\Biggr)\varphi_k(t)&=&0\ ,
\end{eqnarray}
where the effective mass $\cal M$ is defined through the bare fluctuations
$\Sigma_B$ as:
\begin{eqnarray}
{\cal M}(t)^2&=&m^2+\frac\lambda2\phi(t)^2+\frac\lambda2\Sigma(t)\nonumber\\
\Sigma_B(t)&=&\frac1{4\pi^2}\int_0^\Lambda\,
\coth\Biggl(\frac{\beta W_k}2\Biggr)|\varphi_k(t)|^2k^2\,dk\nonumber\\
W_k&=&\sqrt{k^2+{\cal M}(0)^2}\ ,
\end{eqnarray}
where $\Lambda$ denotes the cut-off.
\\
The initial conditions read
\begin{eqnarray}
\varphi_k(0)=\frac1{\sqrt{W_k}}
\nonumber\\
\dot\varphi_k(0)=-i\sqrt{W_k}
\end{eqnarray}
One should notice that since these conditions are smooth functions of $k$,
the mode functions $\varphi_k(t)$ will also have the same property at all time
$t.$
The latter definitions old whenever
the initial system corresponds to a thermal state of the free Hamiltonian,
with a temperature of $\beta^{-1}$.
The zero temperature limit can be obtain letting $\beta$ go to $+\infty$.
Hitherto, it is always possible to write the  equation of motion as:
\begin{eqnarray}
\Biggl(\frac{d^2}{dt^2}+{\cal M}(t)^2\Biggr)\psi_\eta(t)&=&0\nonumber\\
\Biggl(\frac{d^2}{dt^2}+k^2+{\cal M}(t)^2\Biggr)\psi_k(t)&=&0,
\end{eqnarray}
Where the effective mass is written as
\begin{eqnarray}
{\cal M}(t)^2&=&m^2+\psi_\eta(t)^2+S_0(t)\nonumber\\
S_0(t)&=&\int_0^\Lambda |\psi_k(t)|^2\,dk
\end{eqnarray}
To get such an expression it is enough to introduce to the following
notations:
\begin{eqnarray}
\psi_\eta&=&\phi\sqrt{\frac\lambda2}\nonumber\\
\psi_k&=&k\varphi_k \sqrt{\frac\lambda{8\pi^2} \coth\frac{\beta W_k}2}
\end{eqnarray}
It is also convenient to introduce:
\begin{eqnarray}
S&=&\psi_\eta^2+S_0=\psi_\eta^2+\int_0^\Lambda |\psi_k(t)|^2\,dk\nonumber\\
r_k&=&m^2+k^2\nonumber\\
r_\eta&=&m^2
\end{eqnarray}
This quantity S appears as a general quadratic sum, this will play the same
role than the sum $\sum q_i^2$ in the classical analysis. Nevertheless
one should
notice that now the quantities are complex and that the condensate $\phi$
plays a particular role.
\\
With these notations, the natural quantities that generalise those defined
by equation (\ref{Kclassical}) are:
\begin{eqnarray}
\label{Kdef}
K_k&=&\int_0^\Lambda\,
        \frac{|\psi_k\dot\psi_{k'}-\dot\psi_k\psi_{k'}|^2}{{k'}^2-k^2}\,d{k'}
        -\frac1{k^2}\Bigl|\psi_\eta\dot\psi_k-\dot\psi_\eta\psi_k\Bigr|^2
        +2\Bigl(|\dot\psi_k^2|+(m^2+k^2)|\psi_k^2|\Bigr)
        +|\psi_k|^2S
        \nonumber\\
K_\eta&=&\int_0^\Lambda\,
        \frac{|\psi_\eta\dot\psi_k-\dot\psi_\eta\psi_k|^2}{k^2}\,dk
        +2\Bigl(\dot\psi_\eta^2+m^2\psi_\eta^2\Bigr)+\psi_\eta^2S
\end{eqnarray}
\\
In order to manipulate these quantities we have to know whether there are
well defined. Let first study $K_\eta$, the only tricky term is the $1/k^2$
in the integral, but this in fact a false pole since $\psi_k$ contains
$k$ factor. Since $\varphi_k$ is regular $K_\eta$ is well defined.
\\
The same can be said for the $1/k^2$ term of $K_k$. So for $K_k$ the term
that might be ill defined is the first integral which contains
$1/({k'}^2-k^2)$, but since $\psi_k(t)$ is a smooth function of $k$ and $t$, then
then ratio as finite value when $k'$ approaches $k$, hence the integral is
well defined and so is $K_k$.
\\
Now we shall prove that these quantities $K_k$ and $K_\eta$ {\bf do not}
depend on time. This can be directly done using the equation of motion
for $\psi_\eta$ and the $\psi_k$. The main ingredient of the prove is the
use of the following property:
\begin{equation}
\dot S=\psi_\eta\dot\psi_\eta+\int_0^\Lambda\,
        \Bigl(\dot\psi_k\psi_k^*+\psi_k\dot\psi_k^*\Bigr)
\end{equation}
Lets study $K_\eta$ simple algebraic computation using the equation of
motion leads to:
\begin{eqnarray}
\dot K_\eta&=&\int_0^\Lambda\,\frac1{k^2}
        \Bigl(
        (\psi_\eta\ddot\psi_k-\ddot\psi_\eta\psi_k)
        (\psi_\eta\dot\psi_k-\dot\psi_\eta\psi_k)^*
        +{\rm c.c.}\Bigr)\,dk
\nonumber\\
&&      +4\psi_\eta(\ddot\psi_\eta+m^2\psi_\eta)+2\psi_\eta\dot\psi_\eta S
        +\psi_\eta\dot S
\nonumber\\
&=&     \int_0^\Lambda\,\frac1{k^2}
        \Bigl(
        (\psi_\eta(m^2+S)\psi_k-\psi_k(m^2+k^2+S)\psi_\eta)
        (\psi_\eta\dot\psi_k-\dot\psi_\eta\psi_k)^*
        +{\rm c.c.}\Bigr)\,dk
\nonumber\\
&&      -4\psi_\eta(\ddot\psi_\eta+m^2\psi_\eta)+2\psi_\eta\dot\psi_\eta S
        +\psi_\eta\dot S
\nonumber\\
&=&     \psi_\eta\int_0^\Lambda
        \Bigl(\psi_k(\psi_\eta\dot\psi_\eta-\dot\psi_k\psi_\eta)^*
        +{\rm c.c.}\Bigr)\,dk
        -2\psi_\eta\dot\psi_\eta S+\psi_\eta^2\dot S
\nonumber\\
&=&     2\psi\eta\dot\psi_\eta(S-\psi_\eta^2)
        -\psi_\eta^2(\dot S-2\psi_\eta\dot\psi_\eta)
        -2\psi_\eta\dot\psi_\eta S+\psi_\eta^2\dot S
\nonumber\\
\dot K_\eta&=&0
\end{eqnarray}

A similar computation proves that the $K_k$ are also time independent.
\begin{eqnarray}
\dot K_k&=&\int_0^\Lambda\,\frac1{{k'}^2-k^2}
        \Bigl((\psi_k\ddot\psi_{k'}-\ddot\psi_{k'}\psi_k)
        (\psi_k\dot\psi_{k'}-\psi_{k'}\dot\psi_k)^*
        +{\rm c.c.}\Bigr)\,d{k'}
\nonumber\\
&&      \frac1{k^2} \Bigl((\psi_\eta\ddot\psi_k-\ddot\psi_\eta\psi_k)
        (\psi_\eta\dot\psi_k-\psi_\eta\dot\psi_k)^*
        +{\rm c.c.}\Bigr)
\nonumber\\
&&      +2\Bigl(\ddot\psi_k\dot\psi_k^*+(m^2+k^2)\psi_k\dot\psi_k^*
        +{\rm c.c.}\Bigr)
\nonumber\\
&&      +(\psi_k\dot\psi_k^*-\dot\psi_k\psi_k^*)S
        +|\psi_k|^2\dot S
\nonumber\\
&=&     \int_0^\Lambda\,\frac1{{k'}^2-k^2}
        \Bigl(\psi_k\psi_{k'}(k^2-{k'}^2)
        (\psi_k\dot\psi_{k'}-\psi_{k'}\dot\psi_k)^*
        +{\rm c.c.}\Bigr)\,d{k'}
\nonumber\\
&&      \frac1{k^2} \Bigl(\psi_\eta\psi_k(-k^2)
        (\psi_\eta\dot\psi_k-\psi_\eta\dot\psi_k)^*
        +{\rm c.c.}\Bigr)
\nonumber\\
&&      +2(-S\psi_k\dot\psi_k^*+{\rm c.c.})
\nonumber\\
&&      +(\psi_k\dot\psi_k^*-\dot\psi_k\psi_k^*)S
        +|\psi_k|^2\dot S
\nonumber\\
&=&     \int_0^\Lambda\,
        \Bigl(\psi_k\psi_{k'}
        (\psi_{k'}\dot\psi_k-\psi_k\dot\psi_{k'})^*
        +{\rm c.c.}\Bigr)\,d{k'}
\nonumber\\
&&      \psi_\eta\Bigl(\psi_k
        (\dot\psi_\eta\psi_k-\dot\psi_\eta\psi_k)^*
        +{\rm c.c.}\Bigr)
\nonumber\\
&&      +2(-S\psi_k\dot\psi_k^*+{\rm c.c.})
\nonumber\\
&=&     (\psi_k\dot\psi_k^*+{\rm c.c.})(S-\psi_\eta^2)
        -|\psi_k|^2(\dot S-2\psi_\eta\dot\psi_\eta)
\nonumber\\
&&      +\psi_\eta^2(\psi_k\dot\psi_k^*+{\rm c.c.})
        -2\psi_\eta\dot\psi_\eta|\psi_k|^2
\nonumber\\
&&      -S(\psi_k\dot\psi_k^*+{\rm c.c.})
        +\dot S|\psi_k|^2
\nonumber\\
\dot K_k&=&0
\end{eqnarray}
\\
The final step is to express the conserved quantities in term of the
physical modes, that is $\phi$ and $\varphi_k$; we get the following
bare conserved quantities, the first is given from $K_\eta$ and reads 
\begin{equation}
C_{0,B}=\frac12\dot\phi^2+\frac{m_B^2}2\phi^2+\frac{\lambda_B}8\phi^4
+\frac{\lambda_B}{32\pi^2}\int_0^\Lambda\,\coth\frac{\beta W_k}2
        \Biggl[|\phi\dot\varphi_k-\dot\phi\varphi_k|^2
        +k^2|\varphi_k|^2\phi^2\Biggr]\,dk\ ,
\end{equation}
and the other are a set $C(k)_B$ given from the $K_k$:
\begin{eqnarray}
C(k)_B&=&\frac{k^2}2\Biggl(|\dot\varphi_k|^2+(m_B^2+k^2)|\varphi_k|^2\Biggr)
        +\frac{\lambda_B}{8}\Biggl(k^2|\varphi_k|^2\phi^2
                -|\phi\dot\varphi_k-\dot\phi\varphi_k|^2\Biggr)
\nonumber\\
&&      +\frac{\lambda_Bk^2}{32\pi^2}\int_0^\Lambda\,
                k'^2\coth\frac{\beta W_{k'}}2
                \Biggl[
                \frac{|\varphi_k\dot\varphi_{k'}-\dot\varphi_k\varphi_{k'}|^2}{{k'}^2-k^2}
                +|\varphi_k|^2|\varphi_{k'}|^2
                \Biggr]\,dk'\ .
\end{eqnarray}
One should notice, that in the last set of conserved quantities $C(k=0)_B$ is
indeed trivial, since it is expressed in term of a Wronskian ( $\phi$ and
$\varphi_0$ are solution of the same second order differential equation),
\begin{equation}
C(0)_B=-\frac{\lambda_B}{2}|\phi\dot\varphi_0-\dot\phi\varphi_0|^2\ .
\end{equation}

\section{Hierarchy of bare sum-rules}
\label{section_sum}
Although we have already found an infinite number of relations that can be
considered as sum-rules, it is possible to go further and prove the
existence of relations between several integrals over the modes.
\\
The first sum-rule will be given from
\begin{equation}
E_{0,B}=C_{0,B}+\frac1{4\pi^2}\int_0^\Lambda\,C(k)_B 
        \coth\frac{\beta W_k}2\,dk\ ,
\end{equation}
whereas the other will be derivated from
\begin{equation}
E_{n,B}=\frac1{2\pi^2}\int_0^\Lambda\,k^{2n}C(k)_B 
        \coth\frac{\beta W_k}2\,dk\ ,\qquad n\ge1\ .
\end{equation}
All the quantities $E_{n,B}$ are trivially conserved and a constant may be added
without changing this property.
\\
One can easily show that:
\begin{eqnarray}
E_{0,B}&=&\frac12\dot\phi^2+\frac{m_B^2}2\phi^2
        +\frac{\lambda_B}8\Bigl(\phi^4-\Sigma_B^2\Bigr)
        +\frac1{8\pi^2}\int_0^\Lambda\,k^2\coth\frac{\beta W_k}2\Bigl(
                |\dot\varphi_k|^2+\omega_k^2|\varphi_k|^2\Bigr)
                \, dk\ ,
\nonumber\\
\omega_k(t)&=&\sqrt{k^2+{\cal M}(t)^2}\ .
\end{eqnarray}
This shows that $E_{0,B}$ is nothing but the bare energy. This property was
expected since it is a generalisation of equation (\ref{Eclassical}).
\\
The hierarchy of sum-rule will be given from the newly constructed
conserved quantities $E_{n,B}$ for $n$ greater than 1, hereafter we shall use
the following notation to express some integrals over the modes, that are
indeed momenta:
\begin{eqnarray}
\Sigma_{n,B}&=&\frac1{4\pi^2}\int_0^\Lambda\,k^{2n}|\varphi_k|^2
                \coth\frac{\beta W_k}2\, dk
\nonumber\\
\Theta_{n,B}&=&\frac1{4\pi^2}\int_0^\Lambda\,k^{2n}|\dot\varphi_k|^2
                \coth\frac{\beta W_k}2\, dk
\nonumber\\
\Xi_{n,B}&=&\frac1{8\pi^2}\int_0^\Lambda\,k^{2n}
         \Biggl(\dot\varphi_k\varphi_k^*+\varphi_k\dot\varphi_k^*\Biggr)
         \coth\frac{\beta W_k}2\, dk
\end{eqnarray}
One should notice that $\Sigma_{1,B}$ is nothing but$\Sigma_B$.
\\
Using these notations it is straightforward to get
\begin{eqnarray}
E_{n,B}&=&\Theta_{n+1,B}+m_B^2\Sigma_{n+1,B}+\Sigma_{n+2,B}
\\\nonumber&&
        +\frac{\lambda_B}4\Biggl(\phi^2\Sigma_{n+1,B}-\phi^2\Theta_{n,B}
        -\dot\phi^2\Sigma_{n,B}+2\phi\dot\phi\Xi_{n,B}+I_{n,B}
        +\Sigma_B\Sigma_{n+1,B}\Biggr)\ ,
\end{eqnarray}
where $I_{n,B}$ reads:
\begin{equation}
I_{n,B}=\frac1{(4\pi^2)^2}\int_0^\Lambda\,
        |\varphi_k\dot\varphi_{k'}-\dot\varphi_k\varphi_{k'}|^2
        \frac{k^2k'^2k^{2n}}{k'^2-k^2}
                \coth\frac{\beta W_k}2
                \coth\frac{\beta W_{k'}}2\, dkdk'\ .
\end{equation}
It is possible to express $I_{n,B}$ as a sum of integrals
$\Sigma_{p,B}$, $\Theta_{p,B}$ and $\Xi_{p,B}$, to achieve this one should
use the trivial relation:
\begin{eqnarray}
x^n&=&(x-y)Q_n(x,y)+y^n\nonumber\\
Q_n(x,y)&=&\sum_{p=0}^{n-1}x^py^{n-1-p}
\end{eqnarray}
Hence,
\begin{equation}
I_{n,B}=\int_0^\Lambda\,\frac{k^2k'^2dkdk'}{4\pi^2}
        |\varphi_k\dot\varphi_{k'}-\dot\varphi_k\varphi_{k'}|^2
        \Biggl[Q_n(k^2,k'^2)+
        \frac{k'^{2n}}{k'^2-k^2}
        \Biggr]
                \coth\frac{\beta W_k}2
                \coth\frac{\beta W_{k'}}2\ .
\end{equation}
Swapping $k$ and $k'$ leads to:
\begin{equation}
I_{n,B}=\frac1{2(4\pi^2)^2}\int_0^\Lambda\,
        |\varphi_k\dot\varphi_{k'}-\dot\varphi_k\varphi_{k'}|^2
                Q_n(k^2,k'^2)k^2k'^2
                \coth\frac{\beta W_k}2
                \coth\frac{\beta W_{k'}}2\, dkdk'\ .
\end{equation}
Using the following Wronskian relation
\begin{equation}
\label{Wronskian}
\varphi_k\dot\varphi_k^*-\dot\varphi_k^*\varphi_k=2i\ ,
\end{equation}
one gets:
\begin{equation}
\label{WronskianUse}
|\varphi_k\dot\varphi_{k'}-\dot\varphi_k\varphi_{k'}|=
|\dot\varphi_k|^2|\varphi_{k'}|^2+|\dot\varphi_k|^2|\varphi_{k'}|^2
+\frac12
(\dot\varphi_k\varphi_k^*+\varphi_k\dot\varphi_k^*)
(\dot\varphi_{k'}\varphi_{k'}^*+\varphi_{k'}\dot\varphi_{k'}^*)
+2
\end{equation}
The last $2$ will just give an additive constant term to $E_{n,B}$ that can
be dropped out, using the latter relation, one can write $I_{n,B}$ as
\begin{equation}
I_{n,B}=\sum_{p=1}^n\Biggl(
        \Sigma_{p,B}\Theta_{n+1-p,B}-\Xi_{p,B}\Xi_{n+1-p,B}
        \Biggr)
\end{equation}
One finally get an infinite number of relations between the
$\Sigma_{p,B}$, $\Theta_{p,B}$ and $\Xi_{p,B}$:
\begin{eqnarray}
E_{n,B}&=&\Theta_{n+1,B}+m_B^2\Sigma_{n+1,B}+\Sigma_{n+2,B}
\nonumber\\&&
        +\frac{\lambda_B}4\Biggl[\phi^2(\Sigma_{n+1,B}-\Theta_{n,B})
        -\dot\phi^2\Sigma_{n,B}+2\phi\dot\phi\Xi_{n,B}
\nonumber\\&&
        +\Sigma_B\Sigma_{n+1,B}
        +\sum_{p=1}^n\Bigl(
        \Sigma_{p,B}\Theta_{n+1-p,B}-\Xi_{p,B}\Xi_{n+1-p,B}
        \Bigr)
        \Biggr]\qquad, n\ge1\ .
\end{eqnarray}
These equations can indeed be seen as a hierarchy of relations between
integrals over the modes. 

\section{Renormalised conserved quantities}
\label{section_renormalisation}
It is has already been proven
\cite{Cormier:1998wk,Boyanovsky:1996sq,Baacke:1998zy}
that the equation of motion for a quantum
scalar field can be written in term of renormalised quantities. Let recall
here the main results that are given with more details in previous studies.
There are several important points. First that since the short distance
behaviour is not related to the out of equilibrium state of the system, the
coupling constant is renormalised as in usual quantum field theory, that
is:
\begin{eqnarray}
\lambda_B&=&\frac{\lambda_R}{1-\frac{\lambda_R}{16\pi^2}\log\frac\Lambda\mu}
\nonumber\\
m_B^2+\frac{\lambda_B}{16\pi^2}\Lambda^2&=&
        m_R^2\Biggl(1+\frac{\lambda_B}{16\pi^2}\log\frac\Lambda\mu\Biggr)
\end{eqnarray}
Where $\mu$ is the renormalisation scale and $B$ and $R$ subscripts denotes
the bare and renormalised quantities.
\\
The second point is that all the modes are not modified since the effective
mass $\cal M$ is not changed:
\begin{eqnarray}
{\cal M}_B(t)&=&{\cal M}_R(t)\nonumber\\
m_B^2+\frac{\lambda_B}2\phi(t)^2+\frac{\lambda_B}2\Sigma_B(t)&=&
        m_R^2+\frac{\lambda_R}2\phi(t)^2+\frac{\lambda_R}2\Sigma_R(t)
\end{eqnarray}
Using these notations, the renormalised equations remains formally the same
\begin{eqnarray}
&&\Biggl(\frac{d^2}{dt^2}+{\cal M}(t)^2\Biggr)\phi(t)\,=\,0
\nonumber\\
&&\Biggl(\frac{d^2}{dt^2}+k^2+{\cal M}(t)^2\Biggr)\varphi_k(t)\,=\,0
\nonumber\\
&&{\cal M}(t)^2\,=\,m_R^2+\frac{\lambda_R}2\phi(t)^2
        +\frac{\lambda_R}2\Sigma_R(t)\ ,
\end{eqnarray}
just the fluctuations $\Sigma$ are renormalised as:
\begin{eqnarray}
\Sigma_R(t)&=&\frac1{4\pi^2}\int_0^\infty\,
k^2\Biggr[
\coth\Biggl(\frac{\beta W_k}2\Biggr)
                |\varphi_k(t)|^2-\frac1k
                +\frac{\Theta(k-\mu)}{2k^3}{\cal M}(t)^2
        \Biggl]\,dk
\nonumber\\
W_k&=&\sqrt{k^2+{\cal M}(0)^2}
\end{eqnarray}
With the unchanged initial conditions:
\begin{eqnarray}
\varphi_k(0)&=&\frac1{\sqrt{W_k}}
\nonumber\\
\dot\varphi_k(0)&=&-i\sqrt{W_k}
\end{eqnarray}
The last point concerns the behaviour of $|\varphi_k|^2$ and
$|\dot\varphi_k|^2$ for large momenta $k$, that can be found using a WKB
expansion, in term of the effective potential $v(t)$, these read:
\begin{eqnarray}
\label{mode_behaviour}
|\varphi_k|^2&=&\frac1k+\frac v{2k^3}+\frac{3v^2-\ddot v}{8k^5}
        +{\cal O}(\frac1{k^7})
\nonumber\\
|\dot\varphi_k|^2&=&k-\frac v{2k}+\frac{\ddot v-v^2}{8k^3}
        +{\cal O}(\frac1{k^5})
\nonumber\\
v(t)&=&-{\cal M}(t)^2
\end{eqnarray}
Now in order to construct renormalised conserved quantities, we start with the
previously found quantities, which are expressed in terms of the bare
coupling constants and fluctuations.
\\
A direct computation leads to:
\begin{eqnarray}
\biggr(1-\frac{\lambda_R}{16\pi^2}\log\frac\Lambda\mu\Biggl)C_{0,B}
        &=&\frac12\dot\phi^2+\frac{m_R^2}2\phi^2+\frac{\lambda_R}8\phi^4
        -\frac{\lambda_R}{32\pi^2}\int_0^\Lambda\,
         \Biggr[2k\phi^2+\frac{\Theta(k-\mu)}k\dot\phi^2\Biggl]\,dk
\nonumber\\
        &&+\frac{\lambda_R}{32\pi^2}\int_0^\Lambda\,
        \Bigl[|\phi\dot\varphi_k-\dot\phi\varphi_k|^2+k^2\phi^2|\varphi_k|^2
        \Bigr]\coth\frac{\beta W_k}2\,dk
\nonumber\\
        &=&\frac12\dot\phi^2+\frac{m_R^2}2\phi^2+\frac{\lambda_R}8\phi^4
\nonumber\\
        &&+\frac{\lambda_R}{16\pi^2}\int_0^\Lambda\,
        \Bigr[|\phi\dot\varphi_k-\dot\phi\varphi_k|^2+k^2\phi^2|\varphi_k|^2
        \Bigl]\frac{dk}{e^{\beta W_k}-1}
\nonumber\\
        &&+\frac{\lambda_R}{32\pi^2}\phi^2\int_0^\Lambda\,
        \Bigl[|\dot\varphi_k|^2+k^2|\varphi_k|^2-2k\Bigr]\,dk
\nonumber\\
        &&+\frac{\lambda_R}{32\pi^2}\dot\phi^2\int_0^\Lambda\,
        \Bigl[|\varphi_k|^2-\frac{\Theta(k-\mu)}k\Bigr]\,dk
\nonumber\\
        &&-\frac{\lambda_R}{32\pi^2}\phi\dot\phi\int_0^\Lambda\,
        \Bigl[\varphi_k\dot\varphi_k^*+\dot\varphi_k\varphi_k^*\Bigl]\,dk
\end{eqnarray}
The integral has been splited into four parts each of which having, as it
will be shown, a finite limit when $\Lambda$ goes to infinity. 
The explicit temperature dependence has been put into the first integral.
The latter is convergent since the mode functions have, in the worse case,
a polynomial behaviour which is washed out by the exponential factor.
\\
Only the last three integrals have to studied with care. This is done
using equations (\ref{mode_behaviour}), which give the asymptotic behaviour
of the modes for large $k$:
\begin{eqnarray}
|\dot\varphi_k|^2+k^2|\varphi_k|^2-2k&=&\frac{v^2}{4k^3}+{\cal O}(\frac1{k^5})
\nonumber\\
|\varphi_k|^2-\frac1k&=&\frac v{2k^3}+{\cal O}(\frac1{k^5})
\end{eqnarray}
This proves the convergence of the two first integrals, the last one is
analysed computing:
\begin{eqnarray}
|\varphi_k\dot\varphi_k^*|&=&|\varphi_k|^2|\dot\varphi_k|^2
\nonumber\\
&=&\Bigl(1+\frac v{2k^2}+\frac{3v^2-\ddot v}{8k^4}\Bigr)
        \Bigl(1-\frac v{2k^2}+\frac{\ddot v-v^2}{8k^4}\Bigr)
        +{\cal O}(\frac1{k^6})
\nonumber\\
        &=&1+{\cal O}(\frac1{k^6})
\nonumber\\
        &=& ({\cal R}e\,\varphi_k\dot\varphi_k^*)^2
        +({\cal I}m\,\varphi_k\dot\varphi_k^*)^2
\end{eqnarray}
The Wronskian relation (\ref{Wronskian}) can be rewritten as
\begin{equation}
        ({\cal I}m\,\varphi_k\dot\varphi_k^*)=1\ .
\end{equation}
This leads directly to
\begin{equation}
({\cal R}e\,\varphi_k\dot\varphi_k^*)
        =\varphi_k\dot\varphi_k^*+\dot\varphi_k\varphi_k^*
        ={\cal O}(\frac1{k^3})
\end{equation}
which proves the convergence of the last integral.
\\
Hence the renormalised conserved quantity $C_{0,R}$ reads
\begin{eqnarray}
C_{0,R}&=&\frac12\dot\phi^2+\frac{m_R^2}2\phi^2+\frac{\lambda_R}8\phi^4
\nonumber\\
&&        +\frac{\lambda_R}{32\pi^2}\int_0^\infty\,
        \Biggl[\Bigl(
        |\phi\dot\varphi_k-\dot\phi\varphi_k|^2+k^2\phi^2|\varphi_k|^2
        \Bigr)\coth\frac{\beta W_k}2
         -2k\phi^2-\frac{\Theta(k-\mu)}k\dot\phi^2\Biggr]\,dk
\end{eqnarray}
Now it is possible to do about the same work with $C_{n,B}$, a similar
computation leads to:
\begin{eqnarray}
\Bigr(1-\frac{\lambda_R}{16\pi^2}\log\frac\Lambda\mu\Bigl)C(k)_B
&=&\frac{k^2}2\Biggl(|\dot\varphi_k|^2+(m_R^2+k^2)|\varphi_k|^2\Biggr)
        +\frac{\lambda_R}{8}\Biggl(k^2|\varphi_k|^2\phi^2
                -|\phi\dot\varphi_k-\dot\phi\varphi_k|^2\Biggr)
\nonumber\\
&&      +\frac{\lambda_Rk^2}{16\pi^2}\int_0^\Lambda\,
                \frac{k'^2}{e^{\beta W_{k'}}-1}
                \Biggl[ \frac{|\varphi_k\dot\varphi_{k'}
                -\dot\varphi_k\varphi_{k'}|^2}{{k'}^2-k^2}
                +|\varphi_k|^2|\varphi_{k'}|^2
                \Biggr]\,dk'
\nonumber\\
&&      +\frac{\lambda_Rk^2}{32\pi^2}\int_0^\Lambda\,
                \Biggl[k'^2\Bigl(
                \frac{|\varphi_k\dot\varphi_{k'}
                -\dot\varphi_k\varphi_{k'}|^2}{{k'}^2-k^2}
                +|\varphi_k|^2|\varphi_{k'}|^2
                \Bigr)
                -2k'|\varphi_k|^2
                \Biggr]\,dk'
\nonumber\\
&&      -\frac{\lambda_Rk^2}{32\pi^2}\int_0^\Lambda\,
                \frac{\Theta(k´-\mu)}{k'}
                        \Bigl(|\dot\varphi_k|^2+k^2|\varphi_k|^2
                \Bigr)\,dk'
\end{eqnarray}
However, as it shall be shown it is not yet possible to let $\Lambda$ go to
infinity in this last expression.
\\
There are three integrals that may diverge:
\begin{eqnarray}
&&\int^\Lambda\Biggr(
        \frac{k'^2}{k'^2-k^2}|\dot\varphi_{k'}|^2+k'^2|\varphi_{k'}|^2
        -2k'-\frac{k^2}{k'}
        \Biggr)\,dk'
\nonumber\\
&&\int^\Lambda\Biggr(
                \frac{k'^2}{k'^2-k^2}|\varphi_{k'}|^2-\frac1{k'}
        \Biggr)\,dk'
\nonumber\\
&&\int^\Lambda \frac{k'^2}{k'^2-k^2}\Biggr(
               \varphi_k^*\dot\varphi_{k'}^*\dot\varphi_k\varphi_{k'} 
               +\varphi_k\dot\varphi_{k'}\dot\varphi_k^*\varphi_{k'}^* 
        \Biggr)\,dk'
\end{eqnarray}
Since we are interested in the behaviour for large $k'$ the lower bound
of the integral has been omitted, and is indeed assumed to be greater the $k$
and $\mu$, that is avoiding the possible divergences for $k'$ going to $k$
that are known not to exist.
\\
The expansions already used previously proves that the integrants of the to
first integrals are such that
\begin{eqnarray}
\frac{k'^2}{k'^2-k^2}|\dot\varphi_{k'}|^2+k'^2|\varphi_{k'}|^2
        -2k'-\frac{k^2}{k'}
&=&{\cal O}(\frac1{k'^3})
\nonumber\\
\frac{k'^2}{k'^2-k^2}|\varphi_{k'}|^2-\frac1{k'}
&=&{\cal O}(\frac1{k'^3})\ .
\end{eqnarray}
This means that the two first integrals have a finite value for large
$\Lambda$.
\\
For that last integral we shall again use the Wronskian relation
(\ref{Wronskian}) as in equation (\ref{WronskianUse}):
\begin{equation}
\varphi_k^*\dot\varphi_{k'}^*\dot\varphi_k\varphi_{k'} 
+\varphi_k\dot\varphi_{k'}\dot\varphi_k^*\varphi_{k'}^* 
=
2+\frac12
(\dot\varphi_k\varphi_k^*+\varphi_k\dot\varphi_k^*)
(\dot\varphi_{k'}\varphi_{k'}^*+\varphi_{k'}\dot\varphi_{k'}^*)
\end{equation}
So the last equation can be rewritten as:
\begin{eqnarray}
&&\int^\Lambda \frac{k'^2}{k'^2-k^2}\Biggr(
               \varphi_k^*\dot\varphi_{k'}^*\dot\varphi_k\varphi_{k'} 
               +\varphi_k\dot\varphi_{k'}\dot\varphi_k^*\varphi_{k'}^* 
        \Biggr)\,dk'
\nonumber\\
&=&\frac12\int^\Lambda \frac{k'^2}{k'^2-k^2}
        (\dot\varphi_k\varphi_k^*+\varphi_k\dot\varphi_k^*)
        (\dot\varphi_{k'}\varphi_{k'}^*+\varphi_{k'}\dot\varphi_{k'}^*)
        \,dk'
        +\int^\Lambda\frac{2k^2}{k'^2-k^2}\,dk'
        +\int^\Lambda2\,dk'\ ,
\end{eqnarray}
where just the last integral will give rise to a divergence, but this
integral is indeed a constant so can be dropped without adding a time
dependent quantity.
\\
After letting $\Lambda$ go to infinity the renormalised $C(k)_R$ read:
\begin{eqnarray}
C(k)_R
&=&\frac{k^2}2\Biggl(|\dot\varphi_k|^2+(m_R^2+k^2)|\varphi_k|^2\Biggr)
        +\frac{\lambda_R}{8}\Biggl(k^2|\varphi_k|^2\phi^2
                -|\phi\dot\varphi_k-\dot\phi\varphi_k|^2\Biggr)
\nonumber\\
&&      +\frac{\lambda_Rk^2}{32\pi^2}\int_0^\infty\,
                \Biggr[k'^2\coth\frac{\beta W_{k'}}2
                        \Biggr( \frac{|\varphi_k\dot\varphi_{k'}
                        -\dot\varphi_k\varphi_{k'}|^2}{{k'}^2-k^2}
                        +|\varphi_k|^2|\varphi_{k'}|^2\Biggl)
\nonumber\\
&&\qquad      -2 -2k'|\varphi_k|^2
                -\frac{\Theta(k'-\mu)}{k'}\Biggr(
                        |\dot\varphi_k|^2+k^2|\varphi_k|^2\Biggl)
                \Biggr]\,dk'
\end{eqnarray}

\section{Renormalised sum-rules and momenta}
\label{section_renormalised_momenta}

The aim of the present section is to generalises the sum-rules
\begin{eqnarray}
E_{n,B}&=&\Theta_{n+1,B}+m_B^2\Sigma_{n+1,B}+\Sigma_{n+2,B}
\nonumber\\&&
        +\frac{\lambda_B}4\Biggl[\phi^2(\Sigma_{n+1,B}-\Theta_{n,B})
        -\dot\phi^2\Sigma_{n,B}+2\phi\dot\phi\Xi_{n,B}
\nonumber\\&&
        +\Sigma_B\Sigma_{n+1,B}
        +\sum_{p=1}^n\Bigl(
        \Sigma_{p,B}\Theta_{n+1-p,B}-\Xi_{p,B}\Xi_{n+1-p,B}
        \Bigr)
        \Biggr]\qquad, n\ge1\ ,
\end{eqnarray}
with renormalised version of the momenta.\\
The pivotal relations that shall be used are the transcription of the
equation of motion in term of the momenta, those read as:
\begin{eqnarray}
\label{momenta_equation}
&&
\dot\Theta_n+2{\cal M}^2\Xi_n+2\Xi_{n+1}=0
\nonumber\\&&
\dot\Sigma_n=2\Xi_n
\nonumber\\&&
\dot\Xi_n+\Sigma_{n+1}+{\cal M}^2\Sigma_n=\Theta_n
\end{eqnarray}
This set of equations are true for bare quantities and we generalise them
for regularised and renormalised momenta.\\
In order, to simplify the discussion the following notation is introduce:
the sum-rules are written as:
\begin{equation}
E_n=\Theta_{n+1}+\Sigma_{n+2}+{\cal M}^2\Sigma_{n+1}+F_n,\qquad n\ge1
\end{equation}
where $F_n$ is a functional depending only on $\Sigma_k$, $\Theta_k$ and
$\Xi_k$ for $k$ at most equal to $n$; a similar relation is also true for
the energy ($E_0$):
\begin{equation}
2E_0=\Theta_1+\Sigma_2+{\cal M}^2\Sigma_1+F_0,
\end{equation}
where $F_0$ depend on $\Sigma_1$, $\varphi$ and $\dot\varphi$.
\\
The prove is inductive, we shall construct all the renormalised momenta
order per order and thus generalising the sum-rules. Using the standard
expansion (\ref{mode_behaviour}), one is able to renormalise the first
momenta:
\begin{eqnarray}
&&
\Sigma_{0,R}=\frac1{4\pi^2}\int_0^\infty\,\Biggl(
        |\varphi_k|^2\coth\frac{\beta W_k}2-\frac1k\Theta(k-\mu)
        \Biggr)dk\ ,
\nonumber\\&&
\Theta_{0,R}=\frac1{4\pi^2}\int_0^\infty\,\Biggl(
        |\dot\varphi_k|^2\coth\frac{\beta W_k}2-k+\frac vk\Theta(k-\mu)
        \Biggr)dk\ .
\end{eqnarray}
We straightforwardly construct $\Xi_{0,R}$ and $\Sigma_{1,R}$ using the
equation of motion for the momenta:
\begin{eqnarray}
&&\Xi_{0,R}=\frac12\dot\Sigma_{0,R}\ ,
\nonumber\\
&&\Sigma_{1,R}=\Theta_{0,R}-(\Xi_{0.R}+{\cal M}^2\Sigma_{0,R})\ ,
\end{eqnarray}
one finds that
\begin{eqnarray}
&&\Xi_{0,R}=\frac1{8\pi^2}\int_0^\infty\,\Biggl(
        \varphi_k\dot\varphi_k^*\dot\varphi_k\varphi_k^*
        \Biggr)\coth\frac{\beta W_k}dk\ ,
\nonumber\\&&
\Sigma_R=\Sigma_{1,R}=\frac1{4\pi^2}\int_0^\infty\,\Biggl(
        k^2|\varphi_k|^2\coth\frac{\beta W_k}2-k-\frac v{2k}\Theta(k-\mu)
        \Biggr)dk\ .
\end{eqnarray}
One should notice that this last expression is the usual form for the
fluctuations. Whenever one wants to make a finite change in the
renormalisation of the fluctuation $\Sigma$, one should also make a change
in the momenta $\Theta_{0,R}$ and for the renormalised mass $m_R^2$; for
instance a common change is:
\begin{eqnarray}
m_R^2&\rightarrow& m_R^2+\frac{\lambda_R}2\Sigma_{1,R}(0)\ ,
\nonumber\\
\Sigma_{1,R}&\rightarrow& \Sigma_{1,R}-\Sigma_{1,R}(0)\ ,
\nonumber\\
\Theta_{0,R}&\rightarrow& \Theta_{0,R}-\Sigma_{1,R}(0)\ .
\end{eqnarray}
The key point is that this transformation do not modify the equations
of motion (\ref{momenta_equation}).
\\
Using, similar technics one easily finds that:
\begin{equation}
\Xi_{1,R}=\frac12\dot\Sigma_{1,R}=
        \frac1{8\pi^2}\int_0^\infty\,\Biggl(k^2\biggl(
        \varphi_k\dot\varphi_k^*\dot\varphi_k\varphi_k^*
        \biggr)\coth\frac{\beta W_k}2
        +\frac{\dot v}{4k}\Theta(k-\mu)
        \Biggr)dk\ .
\end{equation}
In order to find $\Theta_{1,R}$ and $\Sigma_{2,R}$ one should use the
following relations together
\begin{eqnarray}
&&2E_0=\Theta_1+\Sigma_2+{\cal M}^2\Sigma_1+F_0,\nonumber\\
&&\dot\Xi_1+\Sigma_{2}+{\cal M}^2\Sigma_1=\Theta_1\ .
\end{eqnarray}
Since the energy is a constant it can be renormalised, with a simple
subtraction:
\begin{equation}
E_{0,R}=E_{0,R}(t_0)+E_{0,B}-E_{0,B}(t_0)
\end{equation}
Moreover $F_0$ can be renormalised since it is expressed in term of
renormalisable quantities. So we juste have to solve a set of two linear
equations in $\Theta_{1,R}$ and $\Sigma_{2,R}$.
\\
Now one can generalised this to all the momenta by induction. Let suppose
that one already has renormalised all the momenta up to $\Sigma_{n+1}$,
$\Xi_n$ and $\Theta_n$. On easily get:
\begin{equation}
\Xi_{n+1,R}=\frac12\dot\Sigma_{n+1,R}
\end{equation}
To find  $\Theta_{n+1}$ and $\Sigma_{n+2}$, one uses the fact that $E_n$ is
a constant so can be renormalised, so one have to solve a set of to linear
equations for renormalised quantities.
\begin{eqnarray}
&&E_n=\Theta_{n+1}+\Sigma_{n+1}+{\cal M}^2\Sigma_{n+2}+F_n\ ,\nonumber\\
&&\dot\Xi_{n+1}+\Sigma_{n+2}+{\cal M}^2\Sigma_{n+1}=\Theta_{n+1}\ ,
\end{eqnarray}
where $F_n$ is a function of already renormalised quantities.
\\
Using this method one is able to both get a renormalisation procedure for
the momenta and conserve the form of the sum-rule, that can be written in
terms of renormalised quantities:
\begin{eqnarray}
E_{n,R}&=&\Theta_{n+1,R}+m_R^2\Sigma_{n+1,R}+\Sigma_{n+2,R}
\nonumber\\&&
        +\frac{\lambda_R}4\Biggl[\phi^2(\Sigma_{n+1,R}-\Theta_{n,R})
        -\dot\phi^2\Sigma_{n,R}+2\phi\dot\phi\Xi_{n,R}
\nonumber\\&&
        +\Sigma_R\Sigma_{n+1,R}
        +\sum_{p=1}^n\Bigl(
        \Sigma_{p,R}\Theta_{n+1-p,R}-\Xi_{p,R}\Xi_{n+1-p,R}
        \Bigr)
        \Biggr]\qquad, n\ge1\ ,
\end{eqnarray}

\section*{Outline}

The main result of the paper was to prove the existence of a hierarchy of
sum-rules which were given in term of the mode functions of a scalar field
in out of equilibrium quantum field theory.
\\
The bare and renormalised conserved quantities read
\begin{eqnarray}
C_{0,B}&=&\frac12\dot\phi^2+\frac{m_B^2}2\phi^2+\frac{\lambda_B}8\phi^4
+\frac{\lambda_B}{32\pi^2}\int_0^\Lambda\,\coth\frac{\beta W_k}2
        \Biggl[|\phi\dot\varphi_k-\dot\phi\varphi_k|^2
        +k^2|\varphi_k|^2\phi^2\Biggr]\,dk\ ,
\nonumber\\
C(k)_B&=&\frac{k^2}2\Biggl(|\dot\varphi_k|^2+(m_B^2+k^2)|\varphi_k|^2\Biggr)
        +\frac{\lambda_B}{8}\Biggl(k^2|\varphi_k|^2\phi^2
                -|\phi\dot\varphi_k-\dot\phi\varphi_k|^2\Biggr)
\nonumber\\
&&      +\frac{\lambda_Bk^2}{32\pi^2}\int_0^\Lambda\,
                k'^2\coth\frac{\beta W_{k'}}2
                \Biggl[
                \frac{|\varphi_k\dot\varphi_{k'}-\dot\varphi_k\varphi_{k'}|^2}{{k'}^2-k^2}
                +|\varphi_k|^2|\varphi_{k'}|^2
                \Biggr]\,dk'\ .
\end{eqnarray}
and
\begin{eqnarray}
C_{0,R}&=&\frac12\dot\phi^2+\frac{m_R^2}2\phi^2+\frac{\lambda_R}8\phi^4
\nonumber\\&&
        +\frac{\lambda_R}{32\pi^2}\int_0^\infty\,
        \Biggl[\Bigl(
        |\phi\dot\varphi_k-\dot\phi\varphi_k|^2+k^2\phi^2|\varphi_k|^2
        \Bigr)\coth\frac{\beta W_k}2
         -2k\phi^2-\frac{\Theta(k-\mu)}k\dot\phi^2\Biggr]\,dk
\nonumber\\
C(k)_R
&=&\frac{k^2}2\Biggl(|\dot\varphi_k|^2+(m_R^2+k^2)|\varphi_k|^2\Biggr)
        +\frac{\lambda_R}{8}\Biggl(k^2|\varphi_k|^2\phi^2
                -|\phi\dot\varphi_k-\dot\phi\varphi_k|^2\Biggr)
\nonumber\\
&&      +\frac{\lambda_Rk^2}{32\pi^2}\int_0^\infty\,
                \Biggr[k'^2\coth\frac{\beta W_{k'}}2
                        \Biggr( \frac{|\varphi_k\dot\varphi_{k'}
                        -\dot\varphi_k\varphi_{k'}|^2}{{k'}^2-k^2}
                        +|\varphi_k|^2|\varphi_{k'}|^2\Biggl)
\nonumber\\
&&\qquad      -2 -2k'|\varphi_k|^2
                -\frac{\Theta(k'-\mu)}{k'}\Biggr(
                        |\dot\varphi_k|^2+k^2|\varphi_k|^2\Biggl)
                \Biggr]\,dk'
\end{eqnarray}

\section*{Acknowledgements}

The present work was supported with a grant of ``Graduiertenkolleg
Erzeugung und Zerf\"alle von Elemtarteilchen''. I am specially grateful to
H.~J.~de~Vega for pointing out the existence of the work of
Wojciechowski\cite{SW} and Grosse \cite{Grosse}.

\end{document}